\documentclass[structabstract]{aa}  
\usepackage{natbib}
\usepackage{graphicx}
\usepackage{color}
\usepackage{txfonts}
\def\Rej{\mbox{RE{}\,J~0317-853}}
\def\LB{\mbox{LB{}\,9802}}
\def\HST{\mbox{\it HST}}
\def\Lalpha{\mbox{Lyman\,$\alpha$}}
\newcommand{\Teff}{\hbox{$T_{\rm eff}$}}
\newcommand{\Msolar}{\mbox{\,$\rm M_{\sun}$}} 
\newcommand{\Rsolar}{\mbox{\,$\rm R_{\sun}$}}

\begin{document}
   \title{Constraints on the origin of the massive, hot, and rapidly rotating magnetic white dwarf 
	\Rej\ from an HST parallax measurement}

   \titlerunning{Constraints on the origin of \Rej\ from a parallax measurement}


   \author{B. K\"ulebi
          \inst{1}
          \and
          S. Jordan
          \inst{1}
          \and
         E. Nelan
          \inst{2}
          \and
         U. Bastian 
          \inst{1}
          \and
         M. Altmann 
          \inst{1}
\fnmsep\thanks{Based on observations made with the NASA/ESA Hubble Space Telescope, obtained at the Space Telescope 
Science Institute, which is operated by the Association of Universities for Research in Astronomy, Inc., under NASA 
contract NAS5-26555.  The Guide Star Catalogue-II is a joint project of the Space Telescope Science Institute and 
the Osservatorio Astronomico di Torino.}
}
   \institute{Astronomisches Rechen-Institut, Zentrum f\"ur Astronomie der Universit\"at Heidelberg, M\"onchhofstr. 12-14, D-69120 Heidelberg, Germany\\
              \email{bkulebi@ari.uni-heidelberg.de}
         \and
             Space Telescope Science Institute, 3700 San Martin Dr, Baltimore, MD 21218, USA \\
             }

   \date{Received 18 June 2010/ Accepted 26 July 2010}

 
  \abstract
   {}
   {We use the parallax measurements of \Rej\ to determine its mass, radius, and cooling age and thereby constrain its 
    evolutionary origins.}
   {We observed \Rej\ with the the Hubble Space Telescope's Fine Guidance System to  measure the parallax of \Rej\ and
    its binary companion, the non-magnetic white dwarf \LB. In addition, we acquired spectra of comparison stars with the  
    Boller \&\ Chivens spectrograph of the SMARTS telescope to correct the parallax zero point. For the corrected parallax, 
    we determine the radius, mass, and the cooling age with the help of  evolutionary models from  the literature.}
   {The properties of \Rej\ are constrained using the parallax information. We discuss the different cases of the core 
    composition and the uncertain effective temperature. We confirm that \Rej\ is close to the Chandrasekhar's 
    mass limit in all cases and almost as old as its companion \LB.}
   {The precise evolutionary history of \Rej\ depends on our knowledge of its effective temperature. It is possible that 
    it had a single star progenitor possible if we assume that the effective temperature is at the cooler end of the 
    possible range from $30\,000$ to $50\,000$\,K; if $\Teff$ is instead at the hotter end, a  binary-merger scenario for 
    \Rej\ becomes more plausible.}

   \keywords{Stars: white dwarfs -- stars: magnetic fields -- stars: binaries: close -- stars: distances -- stars: individual: \Rej\ --
stars: individual: \LB
               }

   \maketitle
%

\section{Introduction}
\object{\Rej} is a unique hydrogen-rich white dwarf, which was discovered as an EUV source by the 
ROSAT Wide Field Camera \citep{Barstowetal95}. An analysis of follow-up spectroscopy established 
that the stellar surface is covered by a very strong magnetic field with a range of about 170-660\,MG, 
implying that \Rej\ has one of the strongest magnetic fields detected so far in a white dwarf. 

The optical spectrum together with UV observations taken with  the IUE satellite and the Hubble Space Telescope 
indicated that \Rej\ possesses a very high effective temperatures in the range from 30\,000 to 55\,000\,K;  
\citet{Barstowetal95} achieved their best fit for about 49\,000\,K. A careful analysis of the EUVE spectrum using 
the interstellar medium Lyman lines to account for the interstellar extreme ultraviolet absorption implied  
an effective temperature of $33\,800\,$K \citep{Vennesetal03}. Within these constraints, \Rej\ is one of the hottest 
known magnetic white dwarf (MWD); in any case, it has the highest known temperature of all MWDs with 
a field strength above $20$\,MG \citep{Kawkaetal07,Kulebietal09}.

\citet{Barstowetal95} performed high-speed photometry demonstrating that the optical brightness of \Rej\ varies
almost sinusoidally with a period of $725.4\pm0.9$\,sec and an amplitude of more than $0\fm
1$; these results were confirmed by \citet{Vennesetal03}, who inferred a period of $725.727\pm0.001$\,sec from the 
variation in the circular polarisation. The only reasonable explanation of these results is rotation, implying that \Rej\  
rotates more rapidly than any other known  white dwarf that is not a member of a close binary. The photometric 
variation must be caused by differences in the brightness on various parts of the stellar surface. 
Since no strong absorption lines are detected in the optical, a possible explanation may be a variation 
in the effective temperature over the stellar surface; the reason for this temperature inhomogeneity is currently 
not well understood but is probably  connected to stronger or weaker contributions to the magnetic pressures in 
the stellar atmosphere at different  locations on the stellar surface with different magnetic field strengths.

To achieve a clearer insight into the evolution of \Rej, \citet{Burleighetal99} obtained phase-resolved far-UV 
Hubble Space Telescope (\HST) Faint Object Spectrograph spectra. They found that the previous optical results 
could  generally be confirmed, but that the splitting of the \Lalpha\ component into subcomponents implied 
that the field is probably more complicated than indicated by the mean optical spectrum. By compiling a time series of 
spectra, a model for the magnetic field morphology across the stellar surface was produced using the radiation-transfer 
models through a magnetised stellar atmosphere from \citet[][see paper for a basic description of the methods]{Jordan92} 
and  an automatic least squares procedure. The magnetic geometry could be equally well described by 
an offset magnetic dipole ($x_{\rm off}=0.057$,  $y_{\rm off}=0.004$, and $z_{\rm off}=-0.220$ 
stellar radii), which produces a surface field strength distribution in the range 140-730\,MG or 
an expansion into spherical harmonics up to $l=3$ in which the surface field strengths are constrained to be within 
the range 180-800\,MG .

The mass of the white dwarf was constrained by estimating the absolute magnitudes  (or absolute fluxes) calculated 
from the spectroscopic fit parameters $T_{\rm eff}$, $\log g$ and white dwarf evolutionary models 
\cite[e.g.][]{Wood95,BenvenutoAlthaus99}. The determination of the mass of \Rej\ is, in general not straightforward 
because of the effects of strong magnetic fields; the usual method of using the Stark broadening of  the spectral lines 
to determine $\log g$ and subsequently a mass-radius relation fails in the presence of a magnetic field of 
several hundred MG; the reason is that the standard theory for Stark broadening assumes degenerate  energy levels 
but the magnetic fields help remove this degeneracy.

Nevertheless, the mass determination procedure of \Rej\ can be improved by the knowledge of its distance. \Rej\ 
is inferred to be in a wide-binary double-degenerate system due from its visual companion, which is a non-magnetic 
DA white dwarf companion (\object{\LB}) $7^{\prime\prime}$ away. This object was analysed initially by \citet{Barstowetal95}, 
then later by \citet{Kawkaetal07} (for fit parameters see Table\,\ref{table:LB}). \citet{Barstowetal95} derived 
a distance in the range 33$-$37\,pc with these parameters using the evolutionary models of \citet{Wood92}. 
The physical companionship of \LB\ and \Rej\ has recently been confirmed by \textit{Spitzer} IRAC obsevations \citep{Farihietal08} 
that demonstrated the common proper motion nature of the system.

With an effective temperature of  $50\,000$\,K and assuming a distance of 36\,pc \cite{Barstowetal95} concluded that 
the radius of \Rej\ is about $0.0035$\,\Rsolar\ with a corresponding extreme mass of $1.35$\,\Msolar ($\log g = 9.5$). Later 
\citet{VennesKawka08} derived a mass of $1.32\,\pm\,0.03$\,\Msolar\, using $T_{\rm eff} = 33\,800$\,K, $\log g = 9.4$  and  27 pc 
for the distance. If these conclusions are true, \Rej\ would not only be one of the hottest known MWD but 
also the  most massive ($\approx 1.35$\,\Msolar) isolated  (due to the large separation of \Rej\ and \LB\ we can assume 
that both stars did not interact during stellar evolution) white dwarf discovered so far; only two other white dwarfs 
are known with masses in excess of $1.3$\,\Msolar: LHS\,4033 with a mass in the range 1.31-1.34\,\Msolar \citep{Dahnetal04} 
and the magnetic white dwarf PG\,1658+441 with $1.31\pm 0.02$\,\Msolar\ \citep{Schmidtetal92}.

From the theory of stellar evolution, there are two different ways to produce these massive white dwarfs:
either by single-star evolution of a star with an initial mass higher than 7 or 8\,\Msolar\ \citep{Dobbieetal06,Casewelletal09,Salarisetal09} or 
from the merger of two white dwarfs with  C/O cores \citep[see e.g.][]{Segretainetal97}. The latter scenario is 
supported by the rapid rotation of \Rej.

\citet{JordanBurleigh99} measured the circular polarisation to have a degree of 20\%\ at a wavelength of 5760\,\AA, 
the strongest ever found in a MWD. Together with the assumed small radius and strong gravity in the 
stellar photosphere, this also made \Rej\ a test object for setting limits on gravitational birefringence predicted 
by theories of gravitation, which violate  the Einstein equivalence principle \citep{Preussetal05}.

Since the mass determination of \Rej\ was based entirely on the uncertain spectroscopic distance of the system, 
we applied for observing time with the \HST\ to measure the trigonometric parallaxes of the white dwarf binary 
system to either confirm or disregard the conclusions of \citet{Barstowetal95}. In this paper, we present 
the analysis of the parallax measurement with \HST's Fine Guidance Sensor (FGS).

\begin{table}
\caption{Spectroscopically derived parameters of LB 9802.}             
\label{table:LB}      
\centering          
\begin{tabular}{rrlrl}     
\hline\hline       
\multicolumn{1}{c}{Ref.} & \multicolumn{1}{c}{$V$} & \multicolumn{1}{c}{$T_{\rm eff}$} & \multicolumn{1}{c}{$\log g$}   & \multicolumn{1}{c}{$d_L$} \\
 	 &  \multicolumn{1}{c}{/mag} &  \multicolumn{1}{c}{/K} & &  \multicolumn{1}{c}{/pc} \\
\hline
1 & 14.11 & 16\,030\,$\pm$\,230 & 	8.19\,$\pm$\,0.05	& 33-37  \\
2 & -     & 16\,360\,$\pm$\,80	&	8.41\,$\pm$\,0.02 	& 30 \\ 
3 & 13.90 & 15\,580\,$\pm$\,200	&	8.36\,$\pm$\,0.05 	& 27   \\
\hline
\end{tabular}
\begin{flushleft}
 ${}^1$\citet{Barstowetal95};  ${}^2$\citet{Ferrarioetal97};  ${}^3$ \citet{Kawkaetal07} 
\end{flushleft}
\end{table}


\section{Observation}
\subsection{Observations with the FGS of the \HST}
The observations of the magnetic white dwarf \Rej\ ($\alpha_{\rm ICRS}=03^{\rm h}17^{\rm m} 16\fs1750$, $\delta_{rm ICRS}=-85\degr 32^{\prime} 25\farcs 45$)
and its non-magnetic white dwarf companion \LB\  ($\alpha_{\rm ICRS}=03^{\rm h}17^{\rm m} 19\fs 3050$, $\delta_{rm ICRS}=-85\degr 32^{\prime} 31\farcs 15$)
with the Hubble Space Telescope  were performed with the Fine Guidance Sensor 1r (FGS\,1r) at three epochs (March 2007, September 2007, and March 2008, see Table\,\ref{table:visits}).
The Fine Guidance Sensor is a two-axis, white-light shearing interferometer that measures the angle between a star 
and \HST's optical axis by transferring the star's collimated and compressed light through a polarising beam splitter and a 
pair of orthogonal Koesters prisms \cite[see][for a description of the instrument design]{Nelanetal98, Nelan10}. When FGS\,1r is operated 
as a science instrument, \HST\ pointing is held fixed and stabilized by FGS2 and FGS3 which operate as guiders.

To derive an astrometric solution for position, proper motion, and parallaxes, 
\Rej, \LB, and the reference field stars had to be observed at a minimum of three epochs, preferably at the seasons of
maximum parallax factor to allow us to cleanly separate their parallaxes from their proper motions. These seasons are
separated by about six months.  

\begin{figure}
   \centering
   \includegraphics[width=0.6\textwidth]{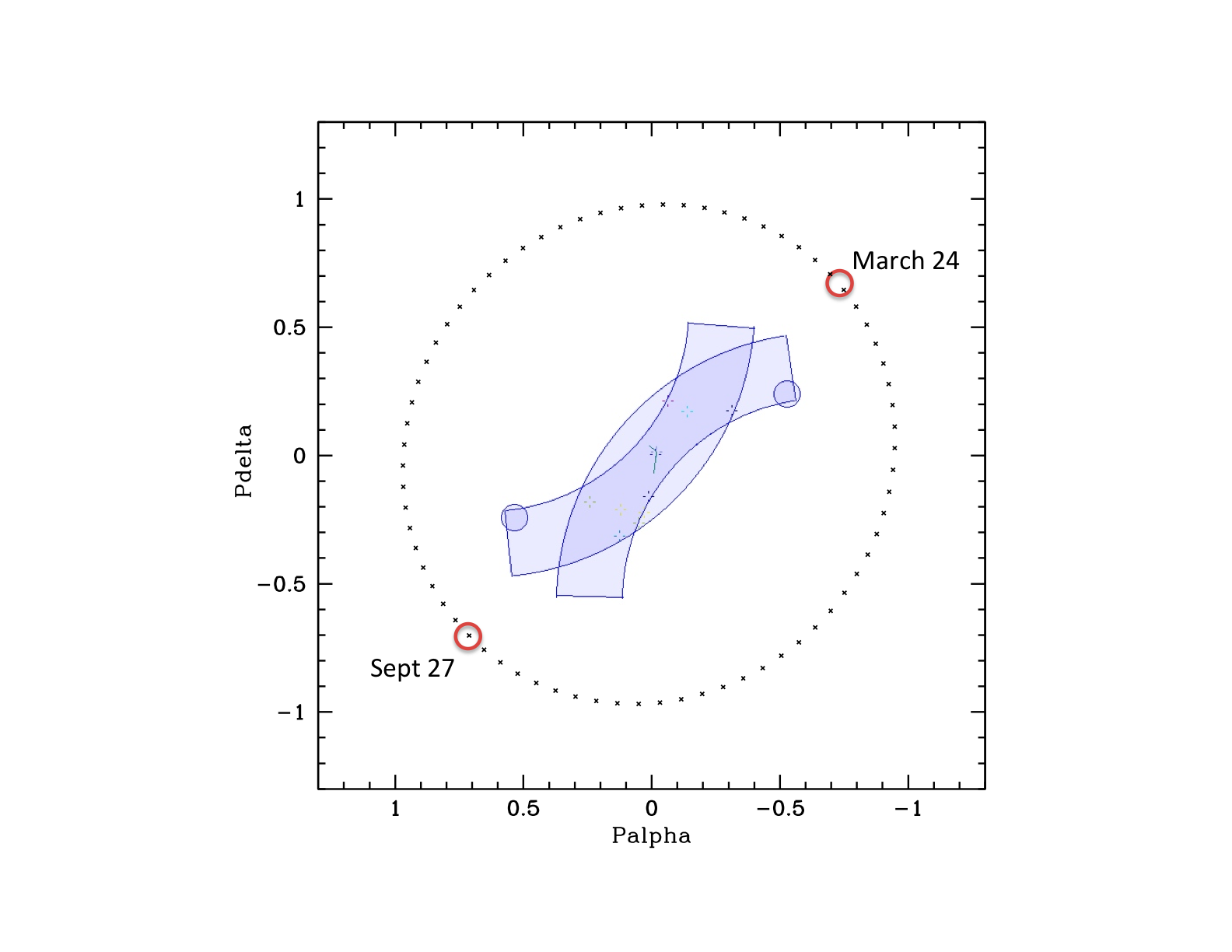}
\caption{The parallactic ellipse of the \Rej\ field and the orientation of the FGS\,1r field of view at the dates of the observations. The X-axis of the
FGS\,1r is nearly parallel to the line connecting the circles that mark the epochs at which the observations were made.}
              \label{fig:parallax}%
    \end{figure}

Fortunately, the epochs of maximum parallax factor also resulted in \HST\ roll angles (which are constrained by date) such that the 
two white dwarf stars and the optimal set of astrometric reference stars could be observed at all epochs. Figure\,\ref{fig:parallax} 
shows the parallactic ellipse and the orientations of the FGS aperture at the times of the observations (there were two March epochs, 2007 and 2008).  
Experience shows that a minimum of two orbits per epoch are required to achieve the highest possible accuracy in the 
final parallaxes. Table\,\ref{table:visits} provides the dates of the six orbits for our \HST\ programme.
 Since the two white dwarfs are only $\approx7''$ apart, we were able to use the same reference stars for the two white dwarfs using
no more \HST\ orbits than would be necessary for a single parallax measurement. Using identical
reference stars also ensured that the parallax difference between the two putative companion stars was measured more
precisely than their absolute parallaxes since the measurements share the same correction of relative to absolute parallax. In addition, their
relative proper motions can be measured to provide an additional check on whether or not the two white dwarfs constitute a bound pair.

\begin{table}
\caption{\HST\ orbits for \HST\ proposal 10930 and 11300.}             
\label{table:visits}      
\centering          
\begin{tabular}{rrrr}     
\hline\hline       
 \multicolumn{1}{c}{proposal ID} & \multicolumn{1}{c}{start time} & \multicolumn{1}{c}{end time}   & \multicolumn{1}{c}{visit} \\
\hline                    
10930 & Mar 24 2007 17:54:01 & Mar 24 2007 18:53:25 & 01 \\
10930 & Mar 24 2007 19:29:48 & Mar 24 2007 20:29:12 & 02 \\
10930 & Sep 27 2007 03:28:55 & Sep 27 2007 04:28:19 & 03 \\
10930 & Sep 29 2007 01:47:19 & Sep 29 2007 02:46:43 & 04  \\
11300 & Mar 29 2008 02:00:30 & Mar 29 2008 02:59:53 & 01\\
11300 & Mar 29 2008 03:36:19 & Mar 29 2008 04:35:42 & 02\\
\hline                  
\end{tabular}
\end{table}

\subsection{Spectroscopy of the astrometric reference stars}

\begin{table*}
\caption{Coordinates and photometry of \Rej, \LB, and the reference stars.}             
\label{table:coordinates}      
\centering          
\begin{tabular}{lrrrrrrrr}     
\hline\hline       
             name   &  $\alpha_{\rm ICRS}$ & $\delta_{\rm ICRS}$ &\multicolumn{1}{c}{$V$} & \multicolumn{1}{c}{$B-V$}  & \multicolumn{1}{c}{$U-B$}  & \multicolumn{1}{c}{$V-R$}  & \multicolumn{1}{c}{$V-I$}  & \multicolumn{1}{c}{GSC2 $F$\footnotemark[2]}  \\
                   &  & & \multicolumn{1}{c}{/mag} & \multicolumn{1}{c}{/mag}  & \multicolumn{1}{c}{/mag}  & \multicolumn{1}{c}{/mag}  & \multicolumn{1}{c}{/mag}  & \multicolumn{1}{c}{mag} \\
\hline                    
\Rej & $03^{\rm h}17^{\rm m} 16\fs1750$ & $-85\degr 32^{\prime} 25\farcs 45$ & $14.90\pm 0.02$ & $-0.16$\footnotemark[1] & $-1.13\footnotemark[1]$ & $0.01$\footnotemark[1] & $-0.11$\footnotemark[1] & $15.09$  \\
\LB &  $03^{\rm h}17^{\rm m} 19\fs 3050$ & $-85\degr 32^{\prime} 31\farcs 15$ & $14.11\pm 0.02$ & $+0.07$\footnotemark[1] & $-0.68$\footnotemark[1] & $-0.06$\footnotemark[1] & $-0.18$\footnotemark[1] & $14.22$ \\
Ref1\footnotemark[3] &$03^{\rm h} 20^{\rm m} 12\fs918 $&$-85\degr 34^{\prime}  56\farcs 175 $& 9.42 & 0.38\\
Ref2\footnotemark[4] &$03^{\rm h} 18^{\rm m} 52\fs01 $&$-85\degr 35^{\prime}  20\farcs 8 $ &  12.27 & 0.50\footnotemark[4] & & &  &12.55\\
Ref3 &$03^{\rm h} 18^{\rm m} 03\fs1 $&$-85\degr 36^{\prime}  02^{\prime\prime} $& 14.60& 1.14\footnotemark[5]\\
Ref6 &$03^{\rm h} 13^{\rm m} 59\fs7 $&$-85\degr 30^{\prime}  16^{\prime\prime} $& 14.00& 1.04\footnotemark[5]\\
Ref7 &$03^{\rm h} 15^{\rm m} 55\fs9 $&$-85\degr 30^{\prime}  20^{\prime\prime} $& 15.00& 0.84\footnotemark[5]\\
Ref8 &$03^{\rm h} 16^{\rm m} 46\fs3 $&$-85\degr 29^{\prime}  48^{\prime\prime} $& 14.37& 1.10\footnotemark[5]\\
Ref9 &$03^{\rm h} 18^{\rm m} 55\fs1 $&$-85\degr 36^{\prime}  42^{\prime\prime} $& 14.36& 0.63\footnotemark[5]\\

\\
\hline                  
\end{tabular}
\begin{flushleft}
 ${}^1$ From \citet{Barstowetal95};
 ${}^2$ http://tdc-www.harvard.edu/catalogs/gsc2.html;
 ${}^3$ =\object{HD 23298}=TYC9495-788-1\citep{Hogetal98};\\
 ${}^4$ =\object{GSC0949500756};
 ${}^5$ theoretical $B-V$ values interpolated for spectral type and MK class, see Table\,\ref{table:comparison}\\
\end{flushleft}
\end{table*}

\begin{table*}
\caption{Results of the analysis of the reference stars ($V$ magnitude, spectral type and luminosity class, absolute magnitude, distance modulus, 
distance, quantifiable distance error (see text), and parallax and its errors (the subscripts ''$-$'' and ''+'' refer to the distances).}             
\label{table:comparison}      
\centering          
\begin{tabular}{rrrrrrrrr}     
\hline\hline       
star &  \multicolumn{1}{c}{V}   & \multicolumn{1}{c}{spectral} &  \multicolumn{1}{c}{$M_V$} & \multicolumn{1}{c}{$m_V-M_V$} & \multicolumn{1}{c}{$B-V$} & 
\multicolumn{1}{c}{$d$} & \multicolumn{1}{c}{$\delta d$} &\multicolumn{1}{c}{$ \pi^{\delta\pi_-}_{\delta\pi_+}$}\\
     &  \multicolumn{1}{c}{/mag}   & \multicolumn{1}{c}{type} &  \multicolumn{1}{c}{/mag} & \multicolumn{1}{c}{/mag} &\multicolumn{1}{c}{/mag}& \multicolumn{1}{c}{/pc} & \multicolumn{1}{c}{/pc} & \multicolumn{1}{c}{/mas}\\
\hline                    
Ref1 &    9.95 &   F3-4V   &3.48  &     6.47& 0.41 &   197 &  24 & $5.08^{+0.69}_{-0.55}$\\
Ref2 &   12.27 &   F7V     &3.95  &     8.32& 0.50 &   461 &  55 & $2.17^{+0.29}_{-0.23}$\\
Ref3 &   14.60 &   K1-2III &0.48  &    14.12& 1.14 &  6668 & 800 & $0.15^{+0.02}_{-0.02}$\\
Ref6 &   14.00 &   K4V     &6.96  &    7.04 & 1.04 &  256  &  31 & $3.91^{+0.53}_{-0.43}$\\
Ref7 &   15.00 &   K0V     &5.98  &    9.02 & 0.84 &  637  &  76 & $1.56^{+0.22}_{-0.16}$\\
Ref8 &   14.37 &   K1III   &0.55  &   13.82 & 1.10 & 5808  & 720 & $0.17^{+0.02}_{-0.02}$\\
Ref9 &   14.36 &   G3-4V   &4.81  &    9.55 & 0.63 &  813  &  98 & $1.23^{+0.23}_{-0.17}$\\
\\
\hline                  
\end{tabular}
\end{table*}

\begin{figure}
   \centering
   \includegraphics[width=0.5\textwidth]{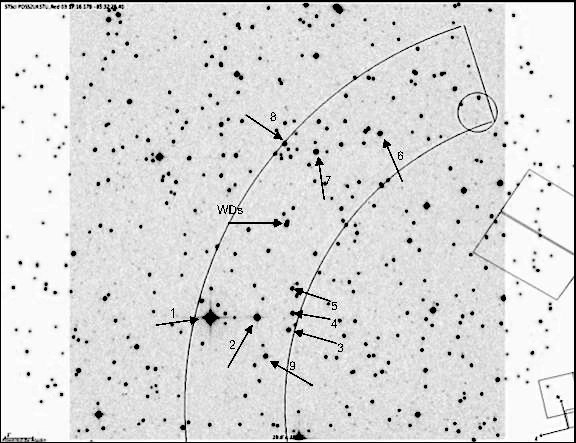}
\caption{The field of the binary WDs \Rej\ and \LB\ and the reference stars Ref1, \dots, Ref9. Ref4 and Ref5 were later omitted
since they were too faint.}
              \label{fig:fc}%
    \end{figure}

Since only relative parallaxes can be measured with \HST, we had to estimate the parallaxes 
of  a sample of reference stars in the vicinity of our target objects which comprise our local reference 
frame (see Figure\,\ref{fig:fc}). Ref4 and Ref5 were not observed by the FGS\,1r since they were not needed.

Spectra of these surrounding stars of similar (or somewhat larger) brightness than \Rej\ were taken in service mode with 
the Boller \&\ Chivens spectrograph of the 1.5m SMARTS telescope,  located on Cerro Tololo at the  
Interamerican Observatory in Chile, in two nights between February 16 and 18, 2008. To ensure 
that the whole optical range is covered, we performed exposures with two gratings (9/Ic and 32/Ib).
Both observing nights were affected by passing clouds and the relatively high airmass ($>1.8$) due to the  large 
declination difference between the observatory's zenith and the target field. Since this could not fully be 
corrected by flux standards, the energy distribution in the blue channel may be compromised.

The classifications for the reference  stars  were performed by comparing the flux-calibrated spectra to 
the templates of \cite{Pickles98}. Since the Pickles library does not cover all spectral subtypes, 
interpolation by eye was performed where appropriate. For late G- and especially K stars, the MK class III 
templates were also looked at because in some cases the star actually turned out to be a giant;  giants 
can be clearly distinguished from dwarfs, which show an indentation at 5200 \AA\ that the giants do not or only slightly 
exhibit. The few  metal-poor and metal-rich templates were also used, although the difference in the Pickles 
spectra is too small to really make a discrimination in this respect.  

\begin{figure}
   \centering
   \includegraphics[width=0.5\textwidth]{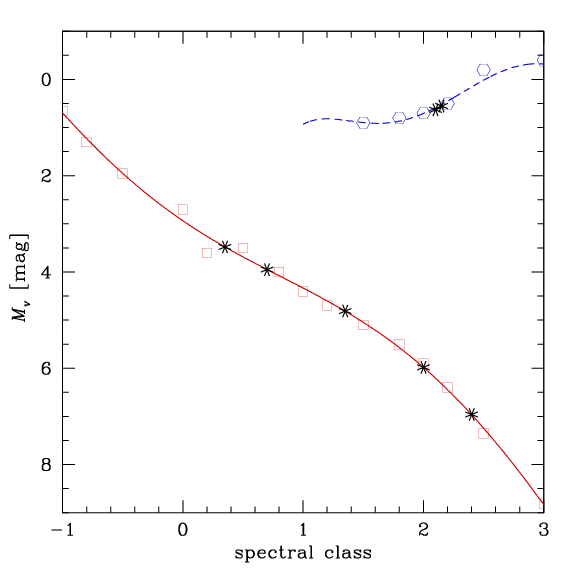}
\caption{The spectrophotometric determinations of the absolute magnitude of the reference stars. The abscissa denotes the 
spectral type encoded in a way that F0 corresponds to 0.0, G0 to 1.0, K0 to 2.0 etc. and the spectral type subdivisions 
being given by the first decimal. The (red) open squares are the loci of main-sequence stars in this HR-diagram, and the 
(blue) open hexagons represent the giants (luminosity class III); the two curves show the resulting fits for both luminosity 
classes. The asterisks show the reference stars of this program. The
spectral (sub)type was determined using low resolution spectra and the absolute magnitude was calculated using the fitted
polynomial.}
              \label{fig:specphot}%
    \end{figure}

The absolute magnitude determination was based on an interpolation of the data taken from \cite{Lang92} 
and Allen's astrophysical quantities \citep{Cox00}. It was achieved by parametrising the spectral type 
so that spectral type F corresponds to 0, G to 1, K to 2, and M to 3, and the spectral type subdivisions 
correspond to the first decimal, i.e. an G2 star would be represented by 1.2. A 5th degree polynomial is then 
fitted to determine the $M_V$ - spectral type relation shown in  Figure\,\ref{fig:specphot}.
This was performed for both luminosity class III and V, assuming that all our stars come 
from these two luminosity classes. The absolute magnitudes of the reference stars were 
then calculated using these two functions with their spectral class parametrised in 
the same way as an argument.

The determination of the errors is not straightforward, since not all error sources can be easily quantified.
The error in the determination of the spectral type was roughly quantified by calculating the absolute magnitude of the
spectral subtypes closest to the determined ones were calculated using the same fit function (for those stars where the 
derived spectral type was in-between two subdivisions, i.e. in the cases of reference stars 1, 3, and 9 the second next
subtype was chosen). The difference between this absolute magnitude and the absolute magnitude obtained for the star is then our
estimate of the error in the absolute magnitude caused by the uncertainty in the  spectral classification. 
This assumes that the error in the spectral type is not  larger than one subdivision, which might not be true in
all cases but should generally be the case. It was generally found that the difference in absolute magnitude between
the measured spectral type and its neighbours is about 0.2 mag, so this 
value was assumed in all subsequent calculations. This error of 0.2 mag corresponds to an error of 12\%\ in distance
(see Table\,\ref{table:comparison}, 7th column).  The corresponding error in the parallax was used for the correction
of the relative parallaxes. Given the relation between parallax and distance, the error in the former is not 
symmetric if that of the former is. The asymmetric nature of the parallax error is represented in column 8 of
Table\,\ref{table:comparison}.  The errors given in Table\,\ref{table:comparison} do not represent the overall error.
A main source of error will most likely be the photometry, which is not of the highest precision. Moreover, our spectra do
not allow us to determine the exact evolutionary status of the objects, which influences the accuracy of the absolute magnitude.
For the same reason, the influence of metallicity cannot be taken into account, and all stars are assumed to be of solar
abundance. Adding these uncertainties with some margin leads to an overall error in distance of 20-30\%, with the
stars Ref7-9 having the larger  errors, since we only have one spectrum (red of Ref7, blue for the other two) for these
objects. Since the parallax is the reciprocal of the distance, the stars with a large distance are the more reliable ones,
especially the two giants (Ref3 and 8).

\begin{table*}
\caption{Astrometric results for \Rej, \LB. and the astrometric reference stars: parallax, proper motion
in right ascension, proper motion in declination, the  standard errors of the proper motion, and the 
standard errors in the fiducial coordinates $\xi$ and $\eta$ of the FGS.
}             
\label{table:results}      
\centering          
\begin{tabular}{rrrrrrrrrl}     
\hline\hline       
            star name &  $\pi$/mas & $\sigma_\pi$/mas 
              & $\mu_\alpha$/mas yr$^{-1}$
              & $\mu_\delta$/mas yr$^{-1}$
              & $\sigma_{\mu_\alpha}$/mas yr$^{-1}$
              & $\sigma_{\mu_\delta}$/mas yr$^{-1}$
             & $\sigma_\xi$/mas
              & $\sigma_\eta$ /mas
              \\
\hline                    
\Rej  &   34.380 &  0.260 & -91.165 & -15.344 & 0.435 &  0.451 &  0.3427 &  0.2085 \\
\LB   &   33.279  & 0.238 &  -78.894 & -27.041 & 0.424 &  0.412 &  0.3042 &  0.2030\\
Ref1  &    4.62  & 0.39 &  10.76 &  19.50 & 0.782  & 0.731 &  0.4544 &  0.1541    \\
Ref6  &    3.51  & 0.40 &   0.00  &  0.00 & 0.000 &  0.000  & 0.4862 &  0.4936    \\
Ref7  &    1.57 &  0.00 & -21.01 &  -8.01 & 0.698 &  0.702 &  0.4552 &  0.3535    \\
Ref8  &    0.17  & 0.00 &   0.00 &   0.00 & 0.000 &  0.000 &  0.4713 &  0.3859  \\
Ref9  &    1.23  & 0.00  &  0.00  &  0.00 & 0.000 &  0.000 &  0.4471 &  0.2317  \\
\hline                  
\end{tabular}
\end{table*}

\section{Analysis of the FGS data}
Our astrometric measurements used FGS\,1r in position mode to observe \Rej, \LB,  and the associated reference field stars.
At each of the three epochs, two \HST\ orbits were used. Within each orbit, FGS\,1r sequentially observed each star several times in a round-robin fashion for 
approximately 30 seconds. The standard FGS data reduction algorithms \citep{NelanMakidon02} were employed to remove instrumental and 
spacecraft artifacts (such as photon shot noise, spacecraft jitter and drift, optical distortion of the FGS, differential velocity aberration, etc).
The calibrated relative positions of the stars in each of the six visits were combined using a six parameter overlapping plate technique that solves for the parallax and proper motion of each star. This process employed the least squares model GaussFit \citep{Jefferysetal88} to find the minimum $\chi^2$ best solution.

The results of the FGS measurements for \Rej, \LB, and the reference stars are given in
Table\,\ref{table:results}.  The $\sigma_\xi$  and $\sigma_\eta$ are the 1\,$\sigma$ errors of the fit of the 
stars onto the ``master plate''. Likewise, the parallax and proper motion errors are the 1\,$\sigma$ dispersion 
in those values measured for the individual observations (e.g. \Rej\ and \LB\ were observed approximately four to 
five times each in every \HST\ orbit, for a total of 24 to 30 individual measurements).
The errors quoted in Table\,\ref{table:results}  are typical of the performance of the FGS\,1r \citep[for comparison see e.g.][]{Benedictetal07},
indicating that our observations are nominal. The best solution was obtained by directly solving for the
trigonometric parallax of Ref1 and Ref6, for which we obtain values consistent with their predicted spectroscopic
parallaxes. Likewise, we derive the optimal solution when we use the FGS\,1r data to solve for the proper motion of 
Ref6 and Ref7. The bulk proper motion of the field is constrained by setting Ref6, Ref8, and Ref9 to have no proper 
motion. The astrometric reference star Ref2 was not used because FGS\,1r resolved it to be a 
wide binary system, which caused an acquisition failure in the second epoch.

The parallaxes for \Rej\ and \LB\  differ by 1.101 mas,  which is about four times the 1\,$\sigma$ 
of their  individual errors. This result includes an application of the  standard ``lateral colour''  
correction that removes the apparent shift of an object's position in the FGS field of view due to the refractive
elements in the instrument's optical train. The correction is given as
$\delta x=(B-V)\cdot lcx$ and
$\delta y=(B-V)\cdot lcy$.
The coefficients $lcx=-1.09$ mas and $lcy = -0.68$ mas are derived from the observed relative positions of
two calibration stars,
LATCOL\_A ($B-V=1.9$),
and
LATCOL\_B ($B-V=0.2$), 
at several \HST\ roll angles. 
However, \Rej\ is significantly hotter and bluer than the blue calibration star LATCOL\_B. It is clear from Figure\,\ref{fig:parallax} that an error in the
lateral colour correction (especially in this case, along the FGS X-axis, which is nearly aligned with the line connecting the two circles marking the dates of the observations) will produce an error in the measured parallax. To evaluate the validity of applying the standard lateral colour correction (which is based solely on a star's value of B-V) to \Rej, we revisited the interpretation of the  astrometric results of the lateral colour calibration observations. Details of this "plausibility" investigation will be published as an STScI FGS Instrument Scientist Report (Nelan, in preparation) but summarized here.

The spectral energy distribution (SED) of the two lateral colour calibrations stars, in addition to \LB, and \Rej\ were convolved with the wavelength-dependent sensitivity of the FGS over its bandpass (the sensitivity decreases from $\approx20\%$ at $4000\AA$ to $\approx2\%$ at $7000 \AA$ in a near linear fashion, where sensitivity refers to the probability that a photon will be detected). The number of photons observed (i.e., actually detected) by the FGS for each star at a given wavelength ($N_{photons}(\lambda)$) was normalized to unity at (for the moment) an arbitrary $\lambda_{o}$. The effective FGS colour of each
star was then defined to be the ratio of the wavelength weighted sum  ($\sum ((\lambda_{o}-\lambda)*N_{photons}(\lambda))$ for all $\lambda<\lambda_{o}$
(the "blue" sum) to the similar "red" sum ($\sum ((\lambda-\lambda_{o})*N_{photons}(\lambda))$ for all $\lambda\geq\lambda_{o}$ over the FGS bandpass.
The value of $\lambda_{o}$ is the boundary between the blue and red such that for a source emitting the same number of photons at every
wavelength the blue and red wavelength weighted sums are equal and the colour ratio is unity. For the FGS, we find that $\lambda_{o}=5092\AA$.

The SEDs of both the red calibration star LATCOL\_A and \Rej\ were represented as black body curves with $T=2\,900$\,K
and $T=50\,000$\,K, respectively, while LATCOL\_B and \LB\ were represented by stellar model atmospheres using a code based upon the Kurutz models.
For LATCOL\_B, a solar abundance, $\Teff=8\,000$\,K, and $\log g=4.1$  were assumed. For \LB, we assumed a hydrogen-atmosphere
white dwarf with $\Teff=16\,030$\,K and $\log g=8.2$. Using these SEDs, we computed for each star the wavelength weighted blue/red ratio
described above, for which we found (blue/red) = 0.13, 1.42, 1.79, and 2.54 for LATCOL\_A, LATCOL\_B, \LB, and \Rej, respectively. (A more
precise estimate of the blue/red ratios for these four stars will use observed SEDs, which are currently unavailable. Here we  simply evaluate the
plausibility of this concept.)

If we assume that the lateral colour shift in the relative position of two stars is proportional to the difference in their blue/red ratios, we can use the 
the astrometric results of the lateral colour calibration, which found that the blue star LATCOL\_B was shifted by -1.87 mas relative to LATCOL\_A,
and their blue/red ratios to determine the proportionality constant $\alpha=-1.85/(1.42-0.13)=-1.44$ mas. Applying this to \Rej\ and \LB, we find the
lateral colour-induced shift in the position of \Rej\ relative to \LB\  to be -1.08 mas. The parallax result cited in Table\,\ref{table:results} already includes a
lateral colour correction of -0.25 mas in the position of \Rej\ relative to \LB\ (based solely upon the (B-V) of each star). This differs by -0.83 mas when
using the difference in their blue/red ratios. If we apply this correction, the parallax difference of the two stars is reduced to 0.27 mas, 
which is $\approx$1\,$\sigma$ of the individual measurements. We conclude that the two white dwarfs have the same parallax, and that this 0.27 mas difference
is caused by the imprecise model SEDs used to construct the blue/red ratios.

The measured parallax of \LB\  is also affected by errors in the lateral colour correction, but to a lesser extent since at (B-V)$=0.07$ it is closer
to the colour of the blue calibration star LATCOL\_B (B-V$=0.2$). Nonetheless, the parallax quoted in Table\,\ref{table:results} 
may be too large by up to 0.4 mas, based on the difference in the predicted relative shift between two stars with (B-V) = 0.2 and 0.07 using the 
standard lateral correction and the blue/red ratio correction. Given the imprecision of the (blue/red) based correction, we take the parallax of \LB\ 
to be $\pi=33.279\pm0.238$ mas using the standard lateral colour correction.

\LB\ is $7^{\prime\prime}$ distant from \Rej\ at a position angle P.A. = $145.856\degr$ as measured by FGS\,1r. From the measured
proper motions (Table\,\ref{table:results}), \LB\ is moving away from \Rej\ at $16.26\pm0.86$ mas yr$^{-1}$ along a position 
angle of $133.62\degr$, which is nearly aligned with the line of sight between the two stars. The computation of the 
proper motions is dominated by the observations from the first and third epochs, which were performed at the same \HST\ 
orientation. Therefore the uncertainty in the lateral colour correction has no effect. At a distance of 30.05 pc (calculated 
from the parallax of \LB), this corresponds to $0.489\pm0.026$\,AU yr$^{-1}$, i.e. $2.33\pm0.12$\,km\, s$^{-1}$. We compare 
this tangential space velocity with an estimated orbital speed. If we assume that this is a bound binary system with 
a separation of $7^{\prime\prime}$ (210 AU at 30.05 pc), and with the total mass ranging from 2.02-2.31 $\Msolar$ 
(see Sect.\,\ref{section:mass}), a circular orbit yields a period of 2004 yr (for the higher mass estimate) to 2143 yr 
(for the lower mass); this corresponds to orbital speeds of 3.12-2.92 km s$^{-1}$ for \LB\ with respect to \Rej. 
These estimates are comparable to the tangential space velocity measured by FGS. This result and the close spatial 
proximity of the two stars supports the conclusion that \LB\ and \Rej\ constitute a bound system. 

Although the FGS photometry shows a peak-to-peak amplitude variation between $V=14.60$ and $V=14.84$ (with $0.01$ accuracy 
estimated using \LB\ as a reference) consistent with the result from \citet{Barstowetal95}, the sampling was not good enough 
to confirm the 725\,second photometric variability quantitatively by means of a Fourier analysis of this sparse data set.

\section{Determination of the stellar parameters}
\subsection{Mass and radius determinations of \Rej\ and \LB\ }
\label{section:mass}

\begin{figure}
   \centering
\begin{tabular}{c}
   \includegraphics[width=0.5\textwidth]{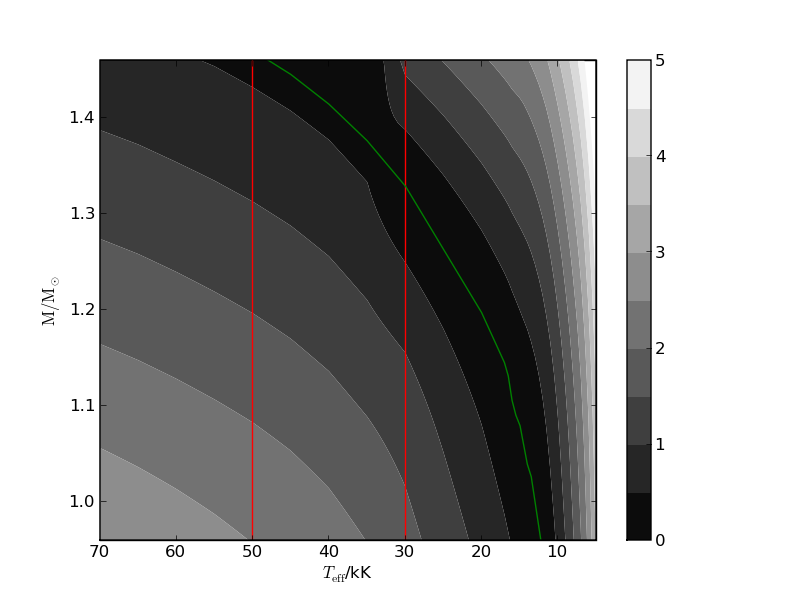}\\
   \includegraphics[width=0.5\textwidth]{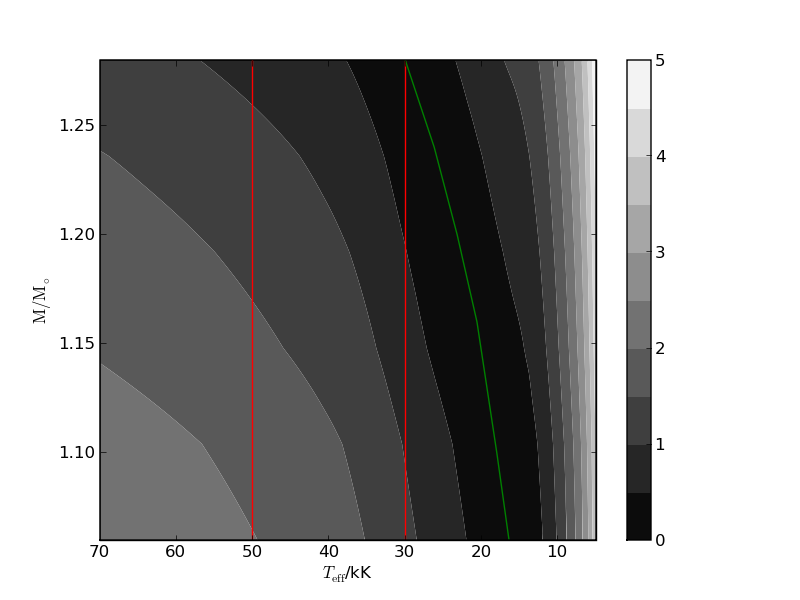}
\end{tabular}
\caption{Contour plots for $|M_V^{\rm obs}-M_V^{\rm theo}|/$mag as a function of mass in $\Msolar$ and $\Teff$ for CO 
(top), ONe (bottom) core compositions constructed according to Eq.\,\ref{eq:MV} and theoretical models from 
\citet{Wood95}, \citet{HolbergBergeron06} for the CO models, and \citet{Althausetal05,Althausetal07} for the ONe models.
The bar to the right indicates the colour coding for the magnitude differences, the line in the darkest 
region $|M_V^{\rm obs}-M_V^{\rm theo}|<0.5$\,mag delinating $M_V^{\rm obs}-M_V^{\rm theo}=0$ and the vertical lines the possible range of effective temperatures 
(30\,000-50\,000\,K). }
    \label{fig:fit}%
\end{figure}

To determine the mass of \Rej, we used synthetic bolometric colours and absolute magnitudes for carbon-oxygen 
(CO) core white-dwarf cooling models with thick hydrogen layers ($M_{\rm H}/M_\ast=10^{-4}$)  
\citep{Wood95,HolbergBergeron06}\footnote{\texttt{http://www.astro.umontreal.ca/$\sim$bergeron/CoolingModels}}; 
when required, we used  oxygen-neon (ONe) core white-dwarf cooling models with hydrogen layers of $M_{\rm H}/M_\ast=10^{-6}$ 
\citep{Althausetal05,Althausetal07}\footnote{\texttt{http://www.fcaglp.unlp.edu.ar/evolgroup/tracks.html}}.

We determined the ``observed'' absolute visual magnitude $M_V^{\rm obs}=V+5\log \pi -5=12.51$ mag from $V=14.90$ 
and $\pi=0.033279^{\prime\prime}$. For a given effective temperature and surface gravity, the theoretical bolometric 
magnitude $M_{\rm bol}$, the bolometric correction B.C.=$M_{\rm bol}-M_V$,  and mass $m$ for \Rej\ were calculated.
The theoretical absolute visual magnitude was defined by 
\begin{equation}
\label{eq:MV}
M_V^{\rm theo}(\Teff,m)=M_{\rm bol}(\Teff,m)-{\rm B.C.}(\Teff,m).
 \end{equation}

The contour plots for  $|M_V^{\rm obs}-M_V^{\rm theo}|$ are shown in Figure\,\ref{fig:fit} for the two possible core compositions. 
For both compositions, a  satisfactory minimum could be reached only for parts of the range of effective temperatures between 30\,000 
and 50\,000\,K because the tables  were limited to an upper value of $\log g=9.5$ for the case of the CO cores ($\log g=9.5$ 
corresponds to a mass of 1.37\,\Msolar\ for 30\,000\,K and a mass of 1.46\,\Msolar\ for 50\,000\,K) and
to an upper limit of 1.28\,\Msolar\ for the ONe models.

We calculated the minimum of $|M_V^{\rm obs}-M_V^{\rm theo}|$ for a given mass of our range  of effective temperatures; when 
a mass solution could not be reached inside the calculated grids, we extrapolated the theoretical magnitudes.

For an effective temperature of $30\,000$\,K, we estimated masses of $1.32\pm0.02$ (CO core) and $1.28\pm0.02$ 
(ONe core). Our CO-core calculations are consistent with the estimates of 1.31-1.37\,\Msolar\ 
\citep{Ferrarioetal97}, who assumed a distance of 30\,pc. The highest temperature for which we could obtain 
a solution in the  $|M_V^{\rm obs}-M_V^{\rm theo}|$ diagram is about $48\,000$\,K from which we inferred 
a mass of 1.46\,\Msolar. Any additional extrapolation may introduce substantial uncertainty because we are then 
approaching the  Chandrasekhar  limit.

\begin{figure}
   \centering
\includegraphics[width=0.5\textwidth]{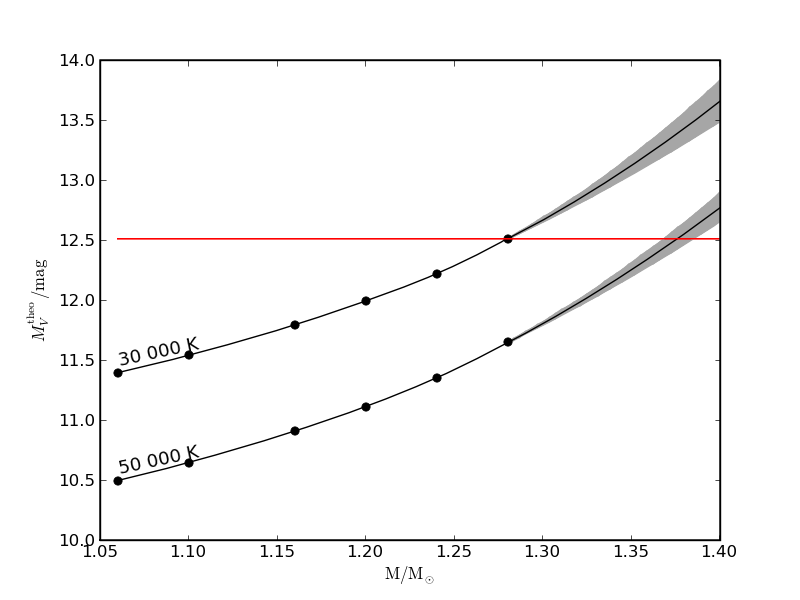}\\
\caption{The mass of \Rej\ versus absolute $V$ magnitudes for an ONe white dwarf. The different curves correspond to the effective 
temperatures 30\,000-50\,000\,K. Above 1.28\,\Msolar, we have to perform an extrapolation for $\Teff>30\,000$\,K.
Since we cannot strictly estimate the extrapolation error, we visually added some uncertainty to the extrapolated values, which
was subsequently used to estimate the errors in Table\,\ref{table:re_mass}. The red line denotes the ``observed'' $M_V$.
}
    \label{fig:extrapolation}
\end{figure}

In the grid of theoretical values for ONe cores, we performed significant extrapolation to obtain 
solutions above 30\,000\,K (see Figure\,\ref{fig:extrapolation}). For $\Teff=30\,000$\,K, we obtained 
a mass of $1.28$\,\Msolar\ and inferred an error of $\pm 0.015$ from the uncertainty in the observed visual 
magnitude and the parallax. For an effective temperature of $50\,000$\,K, we derived $1.38$\,\Msolar\ 
with a slightly higher error estimate of $0.020$\,\Msolar\ due to the uncertainty of the extrapolation.
The results are summarised in Table\,\ref{table:re_mass}.

We applied the same procedure to \LB\ by using our new parallax measurements and the information from the 
literature outlined in Table\,\ref{table:LB}. Our mass estimate for the visual magnitude given by \citet{Barstowetal95}
is consistent with the former results \citep[][see Table\,\ref{table:lb_mass}]{Ferrarioetal97,Kawkaetal07} 
although we find that our calculations with the visual magnitude provided by \citet{Kawkaetal07} 
is incompatible with our mass determination if we assume that the spectroscopically determined masses for \LB\ are correct.

With the knowledge of the $M_{\rm bol}$ for a given mass, the radius can be directly estimated at a given $\Teff$. 
The radius estimates yield slightly different values when two core models are considered (see Table.\,
\ref{table:re_mass}). This is caused assuming the assumption for the hydrogen layer mass to be $M_{\rm H}/M_\ast=10^{-4}$ in 
the CO cooling models \citep{Wood95} versus the $M_{\rm H}/M_\ast=10^{-6}$ content in the ONe cooling models 
\citep{Althausetal05}. This produces different luminosities for a given  effective temperature.

\subsection{Age determination of \Rej\ and \LB}
\label{section:age}
The assessment of the cooling ages of \Rej\ and \LB\ is important to the understanding of the evolutionary history of the system.
It was possible to evaluate the cooling ages of both objects with the mass estimates that we determined. 

\begin{figure}
   \centering
\includegraphics[width=0.5\textwidth]{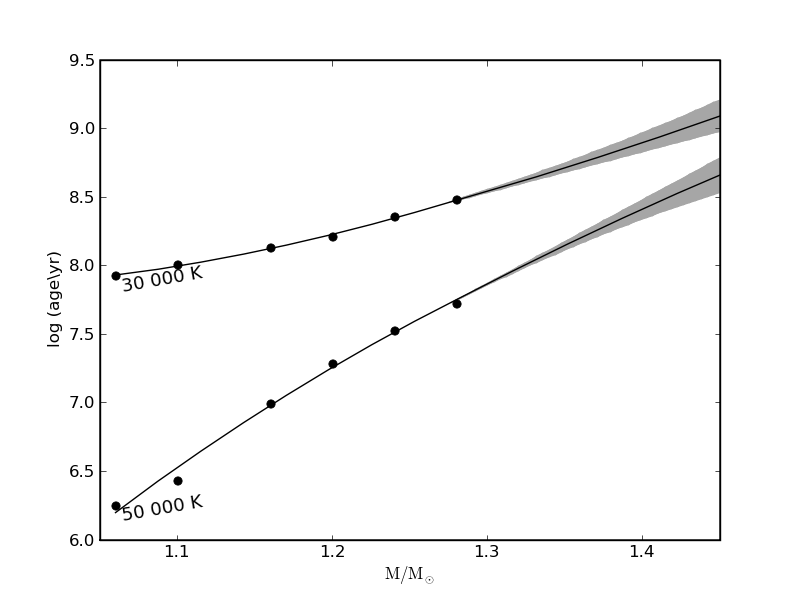}\\
\caption{The mass of \Rej\ versus logarithmic age in years for an ONe core white dwarf. The different curves correspond 
to the effective temperatures 30\,000-50\,000\,K. Since we cannot strictly estimate the extrapolation error we visually 
added some uncertainty to the extrapolated values, which was subsequently used to estimate the errors in Table\,\ref{table:re_mass}.
}
    \label{fig:age_extrapolation}
\end{figure}

For our estimations, we used the grids of white dwarf cooling sequences for CO and ONe cores \citep[e.g.][]{Wood95,BenvenutoAlthaus99}
for their respective range of grid parameters; for masses above the available values, we extrapolated the age values in 
a way similar to that for the visual magnitudes (see Figs.\,\ref{fig:extrapolation} and \ref{fig:age_extrapolation}).

Surprisingly, the difference in the cooling age of the two binary components is smaller than formerly estimated. For both assumed 
chemical compositions, the  cooling age of the non-magnetic white dwarf \LB\ is within the error bars of the cooling age of the 
magnetic and very massive \Rej\ (see Tables\,\ref{table:re_mass} and \ref{table:lb_mass}). For the case of an ONe core  with an 
effective temperature as high as 50\,000\,K, our conclusion is poorly constrained due to the extremely large uncertainties  
introduced by the extrapolation. 
 
Previous age estimates were unreliable because they inferred a cooling age of \Rej\ shorter than that of \LB, simply based on its higher effective 
temperature. If we use the elementary theory of cooling by \citet{Mestel65} assuming for a fixed effective temperature
of the white dwarf, the cooling age is a function of the mass and radius $t_{\rm cool}\propto M/R^2$. This means that the cooling age for 
low-mass white dwarfs ($<0.5\,\Msolar$) is simply proportional to mass  $M^{5/3}$. As the mass of the white dwarf approaches the Chandrasekhar 
limit the radius asymptotically approaches zero, which means that ages for a given effective temperature depend even more strongly on the mass.

The masses estimated here are quite close to the Chandrasekhar limit ($\geq 1.30\,$\Msolar) where post-Newtonian corrections 
should be considered for the stellar equilibrium \citep{Chandrasekhar64,ChandrasekharTooper64}. However, these corrections mostly affect 
the dynamical stability of the star, leading to collapse before reaching the Chandrasekhar limit, but induce only small corrections to 
mass-radius relationship. This is because the estimated radii are three orders of magnitude larger than the Schwarzschild-radius: 
$GM/c^2R_{WD}\sim10^{-3}$. Hence, we do not expect any effect on our mass determinations, as also noted by \citet{KoesterChanmugam90}.

\section{The evolutionary history of the \LB and \Rej\ system}
\label{section:evol}
\begin{table}
\caption{Mass and age estimations for \Rej\ using different core compositions and temperatures. The differences in radius estimates are 
caused by the different hydrogen content for different core models (see Sec.\,\ref{section:mass}).}
\label{table:re_mass}
\centering          
\begin{tabular}{lrlll}     
\hline\hline
Core	& $\Teff$/K 	& mass/$\Msolar$ & radius/$0.01\Rsolar$	&	$t_{\rm cooling}$/Myr \\
\hline
CO  	&	30\,000	&	$1.32\pm0.020$	&  $0.405\pm0.011$	& $281_{-31}^{+36}$ \\
	&	50\,000	&	$>1.46$		&  $0.299\pm0.008$	& $>318$	  \\		 	
ONe 	&	30\,000	&	$1.28\pm0.015$ 	&  $0.416\pm0.011$	& $303_{-38}^{+40}$ \\
	&	50\,000	&	$1.38\pm0.020$ 	&  $0.293\pm0.008$	& $192_{54}^{+110}$ \\
\hline                  
\end{tabular}
\end{table}

\begin{table}
\caption{Mass and age estimations for \LB\ using different $V$ magnitudes in the literature and an average effective 
temperature of 16\,000 K.}             
\label{table:lb_mass}
\centering          
\begin{tabular}{rrr}     
\hline\hline
$V$/mag 	& mass/$\Msolar$ & $t_{\rm cooling}$/Myr \\
\hline
14.11\footnotemark[1] 	&	$0.84\pm0.05$	&	$279_{-39}^{+68}$ \\
13.90\footnotemark[2]	&	$0.76\pm0.05$	&	$223_{-30}^{+36}$ \\
\hline                  
\end{tabular}
\begin{flushleft}
${}^1$ using the visual magnitude from \citet{Barstowetal95}; ${}^2$ using the visual magnitude from \citet{Kawkaetal07}
\end{flushleft}
\end{table}

The projected distance of 210 AU between the two white dwarfs and their small relative proper motion suggest that they are 
companions and therefore share a common origin. The ages of both objects should therefore be equal or comparable within 
the error bars; this condition must be fulfilled for  the correct evolutionary schemes of both white dwarfs. 

The case of \LB\ is straightforward because its evolutionary history  is not complicated by either a strong magnetic field or 
an extreme mass. Therefore, the simple single-star evolution of  \LB\ places constraints on the total age of \Rej.

As mentioned above, previous analyses suggested a younger age for \Rej\ than \LB\ and for this reason the system 
was assumed to have an ``age dilemma''  \citep{Ferrarioetal97}. Therefore, an alternative scenario  was proposed in which 
 \Rej\ is a result of the merging of two  white dwarfs that have lower-mass progenitors. 

\subsection{Single-star origin of \Rej}
\label{sec:single}
With our new results, we undertook a more precise investigation. We firstly considered the single-star scenario for \Rej. 
To determine the total age of \LB\ and \Rej\ from the zero-age main-sequence (ZAMS) to their current stage, we used 
the latest semi-empirical initial-to-final-mass relations (IFMR) \citep{Casewelletal09,Salarisetal09} to estimate their initial masses. 
By considering a diverse range of the theoretical schemes to calculateg the IFMR (metallicity, overshoot parameter, etc.), we 
deduced the progenitor mass of \LB\ to be in the range $4.0-4.5$\,\Msolar. 

For the extremely high (final) mass of \Rej, the corresponding IFMR is quite uncertain. Theoretically, it was shown that 9-10\,\Msolar\ 
mass stars would evolve into massive oxygen-neon (ONe) white dwarfs because of the off-centred carbon ignition in the partially 
degenerate conditions of their cores \citep{Ritossaetal96,Garcia-Berroetal97}. With these constraints in mind, we consider 
more carefully a possible range of initial masses between 8 and 10 solar masses.

The total age (time on the main-sequence plus the  white dwarf cooling time) of \LB\ depends strongly on its initial mass. 
For the  $0.84$\,\Msolar\ mass \LB, the initial masses in the range $4.0-4.5$\,\Msolar\ 
yield main-sequence lifetimes of 170-130\,Myr (the progenitor ages were calculated using the evolutionary tracks from 
\citet{Bertellietal09} for solar metallicity). This means that the total evolutionary age of \LB\ is in the range $410-450$\,Myr.

With 40-30 Myr, the pre-white-dwarf lifetime is extremely short for progenitor masses  in the range between $8$ and $10\,$\Msolar, 
respectively. For an effective temperature of  $30\,000\,$K and our resulting mass of  $1.28-1.32$\,\Msolar\ for \Rej\ 
\citep[which would be the progeny of a  8\,\Msolar\ star, see][]{Casewelletal09,Salarisetal09}, we derive total ages in the range 
$320-340\,$Myr. 

If alternatively we assume an  effective temperature of  $50\,000\,$K for \Rej\ and a CO core, we end up with a total lifetime of $\approx350$\,
Myr; for the ONe core case, the corresponding value would be $\approx220$\,Myr. We reiterate that our estimate 
for the  errors  is rather large in the ONe case at $50\,000\,K$ (see Table\,\ref{table:re_mass}) because of the uncertainties in the 
extrapolation. Hence, omitting the case with ONe core at $50\,000\,$K, we can say that the total age of \Rej\ is in the range $320-350$\,Myr.

There are additional theoretical uncertainties in the IFMR due to magnetism and rapid rotation that should be important for 
an extreme case such as \Rej. The effect of both of these factors  on the IFMR has been the subject of some discussion. \citet{Dominguezetal96} 
argued that rapid rotation has a positive effect on the core growth, such that a rapidly rotating star of mass 6.5\,\Msolar\ 
may produce a white dwarf of mass 1.1-1.4\,\Msolar. Observational evidence of this was found by \citet{Catalanetal08}. 
\Rej\ is the fastest rotating isolated white dwarf and this rotation may be a relic of a rapidly rotating progenitor. 

The assumption of a 6.5\,\Msolar\ mass star as the progenitor does not relieve the ``age dilemma'' considerably since the progenitor 
age for this case is $\sim70$\,Myr, which does not differ much from the 40-30\,Myr estimated for  8-10\,\Msolar\ mass stars. 
\citet{Catalanetal08} also argued that MWDs are relatively more massive than expected on the basis of their inferred progenitors 
via the IFMR of non-magnetic white dwarfs. However, \citet{WickramasingheFerrario05} and \citet{FerrarioWickramasinghe05} both 
concluded based on their population synthesis studies this effect is of only minor importance. Since the effect of rotation 
and magnetism on the evolutionary age is unclear or rather small, we did not consider them in our age estimations. 

Based on these considerations, we conclude that the total age of \LB\ is $410-450$\,Myr at least $\sim100\,$Myr older 
than the respective value for \Rej\ ($320-350$\,Myr). This discrepancy implies that the single-star evolution scenario 
might not be applicable to \Rej. 

However, the mass estimates leading to the cooling ages determined above neglected the influence of magnetism. The magnetic 
nature of \Rej\ is likely to affect the determination of its mass because of the mass-radius relation. 
\citet{OstrikerHartwick68} discussed the effect of magnetism and rapid rotation on white dwarfs. Both magnetism and rotation 
act against the gravitation, causing an extended radius; hence, white dwarfs with strong internal magnetization have larger radii
for a given mass.


To calculate the cooling tracks from synthetic colours and magnitudes of white dwarfs, mass-radius determinations 
are used implicitly. Hence our estimates of the masses and ages are impaired by the lack of mass-radius relations taking 
into account the effect of the magnetic field. For a white dwarf with 1.05\,\Msolar, the radius is increased by a factor 
${\rm e}^{\frac{3}{3-n}\delta}={\rm e}^{3.5\delta}$, where $\delta$ is the ratio of the magnetic energy to the 
gravitational energy of the star, and $n$ is the polytropic index \citep{ShapiroTeukolsky83}. In the case of \Rej, an 
internal  magnetization of $<B>=10^{12}-10^{13}$\,G seems plausible; this would imply that $\delta \approx 0.1$ and therefore 
an increase in the radius by $\sim40\%$. Since \Rej\ has an even higher mass, $n$ is in this case close to 3 and thereby the increase 
in radius for a given mass is even higher.

For an effective temperature of 30\,000\,K, our measured radius is $0.410\times 10^{-2}\,\Rsolar$, whereas for 50\,000\,K  it is 
$0.295\times 10^{-2}\,\Rsolar$. When we correct the influence of the magnetic field on the radius we end up with a 
higher mass than determined in Sect.\,\ref{section:mass}. If \Rej\ were of higher mass the cooling time would increase so that 
the age dilemma no longer exists for the assumption of single-star evolution.

As an initial consideration, cooling ages of $\sim400\,$Myr, which would diminish the age inconsistency, are possible for 
\Rej, if it has a mass of 1.32\,\Msolar\ rather than 1.28\,\Msolar\ (ONe case; 0.04\,\Msolar\ discrepancy), or 1.38\,\Msolar\ rather than 
1.32\,\Msolar\ (CO case; 0.06\,\Msolar\ discrepancy), for an effective temperature of $30\,000$\,K. The corrected radius of 
$R_0=0.32\times 10^{-2}\Rsolar$ implies a mass of 1.38\,\Msolar\ from the mass-radius relationship. This value implies that the 
corrections are plausibly high enough to account for the missing evolutionary age as discussed above.

If we consider $\Teff=50\,000$\,K for \Rej, the mass estimates based purely on the total evolutionary age of the 
system would imply values well above the Chandrasekhar limit. Although it is known that strong internal magnetic field strengths 
also modify the Chandrasekhar limit \citep{OstrikerHartwick68}, it is still difficult to quantitatively assess the masses and their 
effect on cooling ages in this regime.

\subsection{Binary origin of \Rej}
\label{section:binary}
The merger scenario for ultramassive white dwarfs was initially proposed by \citet{Bergeronetal91} for GD\,50, 
\citet{Marshetal97} proposed that this scenario could explain the properties of the hot and massive white dwarf population.
For  \Rej, it was similarly proposed to explain both the high angular momentum and high mass of this star \citep{Ferrarioetal97}.
\citet{Vennesetal03} also suggested that the scenario could produce a strong and non-dipolar magnetic field. 
They argued qualitatively that the high angular momentum is a result of the total orbital momentum of a coalescing binary 
and that the strong non-dipolar magnetic field can be generated by dynamo processes due to the differential rotation 
caused in turn by the merging. 

The type of binary evolution that can lead to a double-degenerate system has been investigated in detail, 
since it represents a channel for producing SN Ia explosions \citep{Webbink84,IbenTutukov84}. In this 
scenario, a binary system consisting of two intermediate-mass stars (5-9\,\,\Msolar) goes through one or two 
phases of a common envelope (CE) and evolves to a  double white dwarf system. If the final double-degenerate 
system has orbital periods in the range between 10\,s and 10\,h, it will lose angular momentum through gravitational 
radiation and merge in less than a Hubble time. The merging process leads to a massive central product with a surrounding 
Keplerian disk. Depending on the total mass of the system, the temperature in the envelope and the accretion to the  
merger product, the system can evolve either to a SN Ia or by an accretion-induced collapse (AIC) to a neutron star. 
When the total mass of the system is insufficient to create the density and the temperature to burn carbon under 
degenerate conditions, the system will end up as an ultra-massive white dwarf.

To  test whether  this scenario is indeed applicable to the case of \Rej, we have to trace back to the point in the stellar 
evolution where the merging could have happened, using the cooling age of \Rej\ and subtracting it from the total evolutionary age 
of \LB. Using this progenitor age estimate and the theoretical constraints from the theory of binary star evolution, we can estimate 
the masses of the possible merging counterparts.

We begin by estimating the mass of the (secondary) binary component that needs longer to become a white dwarf. To obtain a 
lower limit to its mass, we assume the longest time from the main-sequence to the merging process  considering the mass transfer 
episodes predicted by the binary scenario. After both white dwarfs are formed, the time needed for the binary to merge due to gravitational
radiation depends strongly on the orbital parameters and mass of the double-degenerate system. Depending on the properties of the system, 
coalescence can be as fast as 0.1 Myr or as slow as 200 Myr \citep{IbenTutukov84}. To obtain a lower limit to the total evolutionary time
for the system, we neglect the time needed for the double-degenerate system to coalesce. 

\citet{IbenTutukov85} discussed  the evolution of 3 to 12\,\Msolar\ stars that experience two phases of mass transfer. The phase 
of the mass transfer  can take as long or even longer than the time the star spends on  the main-sequence. For a 5 \Msolar\ star, the 
main-sequence phase lasts $\sim$90\,Myr \citep{Bertellietal09}, while in the binary-evolution scenario it takes 140\,Myr from the main 
sequence until the formation of the white dwarf. This means that  230\,Myr are needed for a 5\,\Msolar\ star to evolve into a white dwarf 
rather than the 100 Myr that we assumed for single-star evolution.

The possible cooling ages considered for \LB\ (280\,Myr) and \Rej\ (280 - 320 Myr, when we assume an  effective temperature of about 
30\,000\,K) imply that the maximum time needed for binary evolution is at most the main-sequence age of \LB, which is 130-170\,Myr 
(for 4.0-4.5\,\Msolar). The upper limit of 170\,Myr is comparably short relative to the 230\,Myr of binary evolution time. This 
provides  a lower mass limit for the system. The resulting mass of a white dwarf that is a product of a 5\,\Msolar\ star in this 
binary evolution scheme is 0.752\,\Msolar\ \citep{IbenTutukov85}, which is lighter than inferred from  the IFMRs determined for 
single-star evolution. 

Since the pre-white-dwarf evolution is too long for an initial 5\,\Msolar, star we need a more massive progenitor hence should end up 
with a secondary white dwarf more massive than 0.752\,\Msolar. For the primary star, we assume that it has only a slightly higher mass 
than the secondary to deduce a lower limit to the total coalescing mass.

However, this assumption leads to serious inconsistencies, because the total mass of two components would result in more than  
$1.5\,\Msolar$ being above the masses estimated for \Rej. This lower limit is  also robust  when we consider mass 
loss. Firstly,  smoothed particle hydrodynamic (SPH)  simulations show that only a very small mass loss is expected during merging 
\citep[$\sim10^{-3}\,\Msolar$, see e.g.][]{Loren-Aguilaretal09}, and secondly, we expect almost all of the Keplerian disk to be 
accreted on the merger product. Wind mass-loss from the Keplerian disk is assumed to be lower than 10\%\ of the accretion 
rate \citep{MochkovitchLivio90}; this means that at least 90\% of the disk is expected to be accreted. \citet{Loren-Aguilaretal09} 
also estimate $0.1-0.3\,\Msolar\ $ for the disk masses, which would imply a total mass loss of $\leq 0.01-0.03\,$\Msolar. 

We note that infrared studies have detected possible disks surrounding massive white dwarfs 
\citep{Hansenetal06}. This included \Rej\ for which no convincing evidence of a disk  was found  in the 
\textit{Spitzer} IRAC bands \citep[see also][]{Farihietal08}. If \Rej\ were the product of a merger of two white dwarfs, all of the 
matter from  the Keplerian disk should have been accreted. In this scenario, total mass limits well above the estimated \Rej\ 
mass cannot be avoided. This estimation  eliminates the possibility of a binary origin for \Rej\ with a current effective 
temperature as low as 30\,000\,K.

However, if the total mass of the binary system does not exceed the estimated value for \Rej, 
the time needed for the accretion of all the material from the  disk is much longer than the evolutionary timescale.
The accretion rate is expected to be $\leq 10^{-12}$ \Msolar/yr  for flows with laminar viscosity \citep{Loren-Aguilaretal09}. 
For disk material of $0.1-0.3\,\Msolar$, that its complete accretion time of $1-3 \times 10^{5}\,$Myr
is three orders of magnitude longer than the evolutionary timescale. 

If the binary scenario were correct, the Keplerian disk should have been observed unless the accretion rate of the disk 
was much higher than theoretically predicted. Only accretion rates higher than  $10^{-10}\,$\Msolar/yr would lead to a 
total disappearance of the disk.

When two equal-mass white dwarfs merge, the symmetry of the process leads to a rotating ellipsoidal composed of CO around 
the white dwarf rather than a Keplerian disk. If  \Rej\ were still in the process of accretion we would have observed CO in the 
spectra but this is not the case. On the other hand, if all the material of the surrounding ellipsoid had already been accreted 
(mass-loss can be neglected as discussed above) the mass of \Rej\ would  be higher than observed (above the Chandrasekhar limit). 

We also considered the possible effect on the cooling ages of additional heating of the white dwarf core due to the 
merging process. Recent SPH simulations indicate the possibility of heating to $\sim\,10^9\,K$ in the core 
\citep{YoonLanger05,Loren-Aguilaretal09}. However, because of the $T^{-5/2}$ dependence of the cooling age according 
to the elementary theory \citep{Mestel65}, the effect of this extra heating on the cooling ages 
is expected to be small ($\sim2\,$Myr) and can be neglected.

When we consider the ONe core case for an effective temperature of $50\,000\,$K, leading to  an average cooling age of 
$\sim190\,$Myr, we end up with an upper limit to the evolution time of the secondary  of 220-260\,Myr; the 40\,Myr  
spread in evolutionary time is only due to the uncertainty in the \LB\ progenitor mass (between 4.0-4.5\,\Msolar). 
Our  estimated upper limit to the total age is comparable to the evolutionary timescale of a 5\,\Msolar\ star in a 
binary system as considered above. However, the cooling time estimate for \Rej\ in this case is considerably uncertain 
(see Table\,\ref{table:re_mass}) due to the extrapolation. Within these large error margins, we would in principle be able 
to obtain a sub-Chandrasekhar mass for the merger product, but this process is very unlikely when we consider the time needed 
for the white dwarfs to merge \citep[$10-100\,$Myr][]{IbenTutukov84}. Nevertheless, the possibility of a binary origin 
for an ONe core \Rej\ at $\Teff=50\,000\,$K cannot be entirely excluded.

We note that the effect of magnetic field strength on the structure of the white dwarf, considered in 
Sect.\,\ref{sec:single} is also important to binary evolution. The implementation of this effect 
leads to an inference of slower cooling for \Rej\ as in the single-star scenario. This would yield shorter progenitor 
timescales for a constant evolutionary time, leading to the lower limits on the total mass of the coalescing white 
dwarfs becoming even more massive. This diminishes again the probability of binary evolution for  
$\Teff=30\,000\,$K. However, for $\Teff=50\,000\,$K the uncertainties still permit the possibility of merging. 
Furthermore, the effect of magnetism on the stellar structure ensure that this scenario remains favourable due to the 
higher Chandrasekhar mass limit \citep{OstrikerHartwick68,ShapiroTeukolsky83}.

\section{Discussion and conclusions}

\Rej\ belongs to the very rare population of ultra-massive white dwarfs with masses exceeding 1.1\,\Msolar. The competing 
theoretical explanations of the  origin of these white dwarfs are single-star evolution versus the merging of two degenerate 
stars. Without considering mass-loss during stellar evolution, we have shown that an  upper limit of 1.1\,\Msolar\ for the 
final white dwarf mass would exist for the white dwarfs because of the ignition of  carbon in the core of the progenitor star. 
However, taking into account the effect of mass loss, high-mass ONe-core white dwarfs can be produced \citep[see][for a review]{Weidemann00}. 
Furthermore, it was proposed that even 9 to 10\,\Msolar\ mass stars evolve into ONe core white dwarfs of mass 1.26 and 
1.15 respectively, because of the off-centred carbon ignition in the partial degenerate conditions of their cores 
\citep{Ritossaetal96,Garcia-Berroetal97}.

In the  light of our current results, we have undertaken a more precise investigation of the possible evolutionary 
scenarios for \Rej. We have shown that the cooling ages are almost the same for the  two components. 
The detailed analysis  very much depends on a precise determination of the effective temperature; for $\Teff=30\,000\,$K, 
we can use the calculations by  \citet{Wood95} and \citet{BenvenutoAlthaus99} and conclude that within the limits of the uncertainties 
\Rej\ is at least as old as \LB. For a consistent interpretation of the system, we also have to take into account the time
 scales of the pre-white-dwarf evolution. The more massive progenitor of \Rej\ should evolve more rapidly than the progenitor
 of \LB. Taking this into account, the total age difference between \LB\ and \Rej\ amounts to  $\sim100\,$Myr if single-star 
evolution is considered.

On the other hand, the alternative binary merger scenario proposed by \citet{Ferrarioetal97} and \citet{Vennesetal03} as a 
solution to this age dilemma has severe drawbacks. When the evolutionary timescales are considered, the progenitor age of 
\Rej\ at $\Teff=30\,000\,$K yields lower limits on the mass of the merger product that is considerably higher than its 
estimated mass for all cases. For \Rej, we have large uncertainties in the cooling age estimate only for an effective 
temperature of $50\,000\,$K, so that we cannot fully exclude the binary scenario.

We have also considered the effects of the magnetic fields on both of the scenarios. Magnetic fields cause an increase 
in radius, hence an underestimate of the mass, which would imply longer cooling ages than estimated. For the case of 
$\Teff=30\,000\,$K, the effect of magnetism makes the single-star scenario possible while further eliminating the 
binary merger origin; for the high  $\Teff$ of $50\,000\,$K, even the inclusion of magnetic effects  ensures
that the single-star scenario is possible; the binary scenario remains possible within our large uncertainties.

With our measurement of the parallaxes and relative proper motion of \Rej\ and \LB\ with \HST's FGS, we have established 
that the wide binary system of these two stars is indeed a bound system. We have estimated the masses and ages of \Rej\ 
and \LB\ based on the current white dwarf cooling tracks for different core compositions and hydrogen layer masses. Owing 
to the magnetic nature of this object, the temperature determination of \Rej\ is difficult and should be repeated in the 
future taking into account all available observations and including a more detailed determination of the magnetic field geometry.

For the mass and radius determination, we have considered the highest and lowest possible temperature  and with these 
estimates we have discussed the evolutionary history and the possible origin of \Rej. Our results show that for a 
cooler, less massive  \Rej\ the binary scenario can be excluded within our uncertainties. We also proposed that the 
``age dilemma'' might be solved when the effects of the magnetism on the structure of the white dwarf is considered. 
If \Rej\ were hotter and more massive, then a binary origin scenario would be more plausible.

\begin{acknowledgements}
     This work was supported by the Deutsches Zentrum f\"ur Luft- und Raumfahrt (DLR) under grant 50 OR 0201. B. K\"ulebi 
is a student of International Max Planck School of Astronomy (IMPRS) and a part of Heidelberg Graduate School of 
Fundamental Physics (HGSFP). We would like thank Enrique Garc{\'{\i}}a-Berro for providing us with the data of their 
SPH simulations and the anonymous referee for his valuable suggestions. 
\end{acknowledgements}

\bibliographystyle{aa}
\bibliography{15237.bbl}

\end{document}